\documentstyle[12pt]{article}

\newcommand{\be}{\begin{equation}}
\newcommand{\ee}{\end{equation}}
\newcommand\keV{{\,\sl keV}}
\newcommand\MeV{{\,\sl MeV}}
\newcommand\GeV{{\,\sl GeV}}
\newcommand\Frac[2]{\mbox{$\textstyle{#1\over#2}$}}
\newcommand{\aeff}{\bar a}
\begin{document}
\thispagestyle{empty}
\hfill{Preprint INR -- 0964/97}

\hfill{December 1997}
\vskip 2cm
\begin{center}
{\large \bf Local duality for the $\rho$-meson width revisited.}
\vskip 1cm
{\bf A.A.Pivovarov}\\
{\small {\em Institute for Nuclear Research of the
Russian Academy of Sciences,}}\\
{\small {\em 60th October Anniversary
Pr., 7a, Moscow 117312, Russia.}}
\end{center}

\vskip 1cm

\centerline{\bf Abstract}

\vskip 0.5cm
\parbox{14cm}{\small Local duality estimates for the
$\rho$-meson leptonic width
and the spectrum of radial excitations are updated.
New perturbative corrections are included in the analysis
that gives good agreement with
experimental data on low energy resonances.}
\vskip 1cm

The method of dispersion sum rules
for spectral functions
is a powerful tool for studying
the low
energy properties of the hadronic spectrum.
The method is based on the analyticity properties of the correlators
of the interpolating quark/gluon currents
and asymptotic freedom of QCD. The technique of sum rules was
successfully used for description of hadronic spectrum in different
forms \cite{history}.
Taking into account
condensates of some local operators within the operator product
expansion (OPE) 
allows one to fix the scale of the problem through internal 
quantities of QCD connected with the properties of the vacuum and to 
predict masses of ground states in a number of channels \cite{SVZ}.

Within last decade much
progress has been made in experimental
investigation of low energy
spectrum of hadrons.
Masses of
hadronic resonances and their couplings to quark currents
are now known with better accuracy. The very existence of some
states is now established more reliably while some of entries have
been removed from particle list \cite{PDG}.
Theoretically, further corrections in the strong coupling constant 
expansion have been obtained for perturbative parts of current 
correlators with a variety of quantum numbers that calls for updating 
an old analysis at the higher level of precision.

While the global duality assumption within OPE is well suited for
description of the ground state resonances, more detailed information
on the spectrum can be obtained by assuming the validity of
the local duality approximation.
The accuracy control is less direct
in this case. In fact, only
an agreement with data justifies
the use of the approximation,
i.e. that the length of the averaging interval is sufficient
for comparison between quark-gluon and hadronic quantities.
Local duality estimates of the characteristics of mesonic spectrum
happened to be rather successful both for light and heavy quarks
\cite{FESR}.
Now corrections of higher order in strong coupling expansion for the
correlation function of the current of massless quarks are available
that makes possible to improve on the accuracy of theoretical
predictions.

In the present note we consider local duality prediction for the
leptonic width
of the $\rho$-meson and masses of its excitations.

The interpolating current for the $\rho$-meson has the
form
\[
j_\mu(x)=\frac{1}{2}(\bar u\gamma_\mu u -\bar d\gamma_\mu d)(x)
\]
which is a neutral component of the isotopic triplet
($J^{PC}=1^{--})$ and is chosen because the resonance residue to the
current can be expressed through the leptonic width
$\Gamma(\rho\rightarrow e^+e^-)$ measured experimentally.

The corresponding correlation function reads
\[
\Pi_{\mu\nu}(q)=i\int \langle Tj_\mu(x)j_\nu(0)\rangle e^{iqx} dx
=(q_\mu q_\nu-g_{\mu\nu}q^2)\Pi(q^2).
\]
It is transverse because of conservation of the current.
We also consider quarks u and d to be massless.
For the Adler's function that is free of overall
subtraction constant one has
\[
D(Q^2)=-Q^2{d\over d Q^2} \Pi(Q^2), \qquad -q^2=Q^2.
\]
The operator product expansion for $D(Q^2)$
with contribution of the gluon vacuum condensate reads
\cite{SVZ}
\be
D(Q^2)=D^{PT}(Q^2)+
{\langle {\alpha_s\over \pi} G^2\rangle}{1\over 12 Q^4}+\ldots
\label{df}
\ee
where
\[
D^{PT}(Q^2)={1\over 8\pi^2}(1+a(Q)+ k_1 a(Q)^2+k_2 a(Q)^3+...)
\]
is the perturbation theory part of the correlator and
\[
a(Q)={\alpha_s(Q)\over \pi}
\]
is the coupling constant of strong interaction.
Here $k_{1,2}$ are known coefficients.

The function $D(Q^2)$ is represented by the dispersion relation
with the spectral density $\rho(s)$
\[
D(Q^2)=Q^2\int_0^\infty {\rho(s)ds\over (s+Q^2)^2}
\]
The perturbative part of the $D$-function gives
the following expression for the spectral density
\[
\rho^{PT}(s) ={1\over 8\pi^2}\left(1+a(\mu^2)+(k_1+\beta_0L)a^2(\mu^2)
\right.
\]
\[
\left.
+\Big(k_2-\Frac13\pi^2\beta_0^2+(2\beta_0k_1+\beta_1)L
+\beta_0^2L^2\Big)a^3(\mu^2)+\ldots\right)
\]
with $L=\ln(\mu^2/s)$, $\beta_{0,1}$ are coefficients of the
$\beta$-function.  The term $-\pi^2\beta_0^2a^3(\mu^2)/3$ stems from
analytical continuation in time-like region of momenta. Such terms can
be resummed in all orders of $a(\mu^2)$ for any given order of the
$\beta$-function \cite{pivanal,groote}.

For fixing normalization we give
the definition of the $\beta$ function
\begin{equation}
\mu^2\frac{da}{d\mu^2}=\beta(a)=-\beta_0
a^2-\beta_1 a^3-\beta_2 a^4-\ldots
\end{equation}
with
\[
\beta_0=\frac94, \qquad \beta_1=4,\qquad\beta_2=\frac{3863}{384}
\]
and numerical values of coefficients $k_{1,2}$
\[
k_1=\frac{299}{24}-9\zeta(3)\approx 1.64,
\]
\[
k_2=\frac{58057}{288}-\frac{779}4\zeta(3)+\frac{75}2\zeta(5)\approx 6.37
\]
in the $\overline{\rm MS}$
scheme for $N_c=n_f=3$.
The coefficient $k_2$ has been recently computed independently
in \cite{Chet} that confirmed the old result \cite{Gorishny}.

Dispersion sum rules for the spectral density can be based on global
or local duality. Global duality uses the procedure of smearing for
the entire positive semiaxis while the assumption of the local duality
is that the smearing can be done for parts of this semiaxis.
Thus the local duality assumption leads to the finite energy 
sum rules (FESR) of the 
form 
\begin{equation}
\int_{s_1}^{s_2} \rho^{ph}(s)s^k ds =\int_{s_1}^{s_2} 
\rho^{th}(s)s^k ds 
\label{fesr0}
\end{equation}
for some integer $k$ and the integral is 
performed around the single resonance, $\rho^{ph}(s)$ and 
$\rho^{th}(s)$ being the physical and theoretical spectral densities 
correspondingly. Powers of energy ($s^k$) can be substituted with more 
general functions with proper analytical behavior.
The actual accuracy of the relation (\ref{fesr0}) depends on the shape
of 
the weight functions. Functions without sharp changes are supposed to 
give the best results and we use $k=0,1$. For intervals not including 
the origin the set $k=-1,0$ is also used.

Perturbative analysis
(without vacuum condensates) of the $\rho$ meson spectrum
within the local duality approach was performed in
\cite{FESR} with the result $s_0=2 m_\rho^2$ and in the leading order
in $\alpha_s$
\begin{equation}
\Gamma^0(\rho\rightarrow e^+e^-)={\alpha^2 m_\rho\over
3\pi} =4.4 \keV
\label{gamma0}
\end{equation}
where $\alpha$ is the fine structure constant,
$\alpha=1/137$, and
$m_\rho=768.5\pm 0.6 \MeV$ \cite{PDG}.
This
result can partly be improved by using perturbative corrections.  With
strong interaction corrections included
in the leading order, the old result of
\cite{FESR} is
\begin{equation}
\Gamma(\rho\rightarrow e^+e^-)={\alpha^2 m_\rho\over
3\pi} (1+\frac{\alpha_s(2 m_\rho^2)}{\pi}) =4.8 \keV
\label{oldgamma1}
\end{equation}
with
$\Lambda=100 \MeV$ that was advocated in \cite{SVZ}.
The estimate (\ref{oldgamma1}) is considerably 
smaller than 
the present experimental value of the leptonic width
\cite{PDG}
\begin{equation} 
\Gamma_{e^+e^-}^{exp} = 6.77\pm 0.32 \keV. 
\label{exper}
\end{equation}

In the present paper we correct the result (\ref{oldgamma1})
adding the nonperturbative corrections 
(the contribution of gluon condensate) and taking into account 
new terms of $\alpha_s$ expansion for the coefficient function of 
the unity operator in eq. (\ref{df}). 
At the level of precision of 5-10\% 
the running of the electromagnetic fine structure constant ($\alpha$) 
that enters the definition of the leptonic width eq. (\ref{gamma0})
through the 
residue of the resonance with the quark electromagnetic
current to the energy scale of 
the order of $1~GeV$ is also important.  All these corrections move the 
result (\ref{oldgamma1}) in the right direction and give accurate 
theoretical prediction in agreement with experimental data on leptonic 
width and masses of radial excitations.

Nonperturbative contribution due to nonzero vacuum condensates can be
easily taken into account \cite{ZFESR}.
For the ground state one has
the system of equations
for $k=0,1$
\begin{eqnarray}
F&=&s_0(1+\aeff(m_\rho^2))\nonumber\\
F m_\rho^2&=&{s_0^2\over 2}(1+\aeff(m_\rho^2))-A
\label{sr}
\end{eqnarray}
with the physical spectral density
of the form of a narrow resonance
approximation
\[
\rho^{ph}(s)=F \delta(s-m_\rho^2).
\]
The parameter $A$
\[
A={\pi^2\over 3}\langle {\alpha_s\over \pi} G^2\rangle = 0.04 \GeV^4
\]
is a contribution of the gluon vacuum condensate
where the standard numerical value
$
\langle {\alpha_s\over \pi} G^2\rangle = 0.012 \GeV^4
$
is used \cite{SVZ}.

The quantity $\bar a$ is an effective charge for the moments.
To make the running coupling constant integrable at small momenta
we use the technique
of effective charges \cite{Gru}. Then in the third order of PT for
the $\beta$ function there is an 
infrared fixed point \cite{MatSte}. The results of
integration are
\[
\int_0^{s_0}ds \bar a(s)=\bar a_0 s_0, \qquad
\int_0^{s_0}ds s \bar a(s)=\bar a_1 \frac{s_0^2}{2}
\]
with $\bar a_0=0.302$, $\bar a_1=0.281$ for $s_0\approx 1.3 \GeV^2$ and
and $\bar a(m_\tau^2)=0.12(0.380/\pi)$ \cite{renRS}
chosen as an input, $m_\tau=1.777 \GeV$ is the $\tau$ lepton mass.
Note that the semileptonic width of the  $\tau$ lepton that is now one
of the best sources for low energy determination of the coupling
constant is given by a particular FESR \cite{BraNarPich}.
In the following we neglect the difference between
$\bar a_0$ and $\bar a_1$ putting $\bar a_0\approx\bar a_1\approx \bar
a=0.3$.

Because both $A$ and $\bar a$ are small we treat them as corrections
and limit ourselves to the linear approximation in these parameters.

From eqs. (\ref{sr}) one has
\[
s_0=2 m_\rho^2 + {A\over m_\rho^2}
\]
and
\[
\Gamma_{e^+e^-} = \Gamma^0(\rho\rightarrow e^+e^-)
(1+\bar a)
(1+{A\over 2m_\rho^4})
=4.4*1.3*1.06= 6.06 \keV
\]
that is still smaller than the experimental number (\ref{exper}).
First factor stems from perturbative corrections and second from
gluonic condensate.

An additional contribution comes from remormalization of the
electromagnetic coupling constant
(for details, see e.g. \cite{EidJeg}).
\[
\bar \alpha(s)= { \alpha\over 1-\Delta \alpha(s)},\qquad
\Delta \alpha(s)= {\alpha\over 3\pi}
\sum_f Q_f^2 N_{cf}\left(\ln{s\over m_f^2}-\frac{5}{3}\right)
\]
where the sum runs over the fermions with masses smaller than $s$,
$Q_f$ is the fermion electric charge,  $N_{cf}=3$ for quarks and
$N_{cf}=1$
for leptons.
We find
\[
\bar \alpha(1.3\GeV^2)= { \alpha\over 1-\Delta \alpha(1.3\GeV^2)}
= {\alpha\over 1-0.025}
\]
that gives another five percent in the width.
Finally we have
\be
\Gamma(\rho\rightarrow e^+e^-)
= 6.37 \keV
\label{res}
\ee
which is the main result of the paper
that is much closer to the experimental value 
(\ref{exper}) though is still somewhat low.
The agreement can be achieved by increasing the numerical value 
of the gluon condensate that is not very plausible because the masses
of excitations will be larger.

The empirical estimate \cite{FESR}
\be
s_0={m_\rho^2 + m_{\rho'}^2\over 2}=1.35 \GeV^2
\label{emp}
\ee
also gives a good agreement with the experimental value, namely
\be
\Gamma^{emp}(\rho\rightarrow e^+e^-)
= 6.76 \keV
\label{resemp}
\ee
because the duality interval (\ref{emp})
is somewhat larger.

Next, the radial excitations can be predicted then using 
the assumption that
the integration border lies exactly in the middle between subsequent
resonances \cite{FESR}.

Corresponding parameters for
$\rho'$ are
$m_{\rho'}^2=1.96 \GeV^2$
($m_{\rho'}=1.40 \GeV)$
while the experimental number is
\[
m_{\rho'}=1449\pm 8 \MeV
\]
with the total width
$
\Gamma=310\pm 60 \MeV$.

It is reasonably good numerically though is
not very well justified just because of large width of the
corresponding state. The point is that the very definition of mass of
the resonance depends on the procedure used for fitting data.
For instance, there is a difference whether the common Breit-Wigner
resonance curve is used in terms of a center mass energy $E$
or the energy square $s=E^2$ (see a thorough discussion in
\cite{EidJeg}).

Even higher excitation is in a reasonable agreement with our naive
formula, namely
\[
m_{\rho(1700)}=1717\pm 13 \MeV
\]
while we predict
\[
m_{\rho(3)}=1.80 \GeV.
\]
For the next one we have
$
m_{\rho(4)}=2.13 \GeV
$
while the experimental number is
\[
m_{\rho(2150)}=2149\pm 17 \MeV
\]

These local duality estimates of the spectrum that are essentially the
linear model of ref. \cite{FESR} are in better agreement with
experiment because experimental situation changed. The state
$\rho(1250)$ moved to
$\rho(1450)$ and $\rho(1600)$ is shifted to $\rho(1700)$($m=1717\pm 13\MeV$)
improving agreement with local duality predictions.
Still for the $\rho(1700)$
the discrepancy is rather large.

Note that for exited states
one can compute the resonance 
parameters using two sets of sum rules for
$k=0,1$ and $k=-1,0$.  The difference between predictions can be
considered as a practical estimate of the accuracy of the method.
Roughly, numerical estimate is needed in any particular case but for
two
sets mentioned above the difference is within 15\%.

Main ingredients of the improvement of the agreement
of finite energy sum rules predictions with experimental data are:
\begin{itemize}
\item
Better experimental data, especially for first two excitations.
The state
$\rho(1250)$ moved to
$\rho(1450)$
that allowed to increase the duality interval for the $\rho$-meson
without contradiction with the spectrum.
\item
Better knowledge of the QCD coupling and understanding
of its infrared behavior
that gives the agreement for the width but practically does not affect
the ratio of the moments for determination of the duality interval
and the mass of the first excitation
\item
An account for the gluonic condensate that moves
numerical estimates in the right direction. Its effect is rather small
for the standard numerical value of the condensate. However
it improves on the agreement.
Note that it cannot be increased much if one would like to preserve
the spectrum of radial excitations within
this approach: for larger condensate width fits
better the experimental value
but predictions for masses move too far.
One should not be too strict on that however because for higher
states the approximation of narrow resonances may become inadequate
and finite width
of resonances should be taken into account.
\item
Renormalization of the fine structure constant also improves the
prediction for the $\rho$-meson leptonic width.
\end{itemize}

Note that perturbative corrections cancel in the ratio for
determination of the duality interval through the mass of the ground
state.
Thus the prediction for the mass of radial excitations is not sensitive to
pQCD corrections while the residue is directly proportional to their
magnitude.
Their account improves the agreement with experimental data.

Agreement with experiment is rather good.
That makes FESR a powerful tool for working out the
phenomenology though the internal mechanism for
determination of the ground state mass is absent and
the mass of the ground state sets the scale in determination of the
spectrum.
Thus these two approaches are complementary.
As far as condensates are used to find a mass of the ground state the
FESR
can be used for further investigation of the spectrum.
Power corrections do not affect higher radial excitations 
within the local
duality approach.

The low energy states are now much better known that allows
to reanalyzed some FESR results with other quantum numbers as well.

{\bf Acknowledgments}

This work  is supported in part by Russian Foundation
for Basic Research
grants Nos. 97-02-17065 and 96-01-01860.

\end{document}